\RequirePackage{fix-cm}
\documentclass[natbib,smallextended]{svjour3}       
\smartqed  
\usepackage{graphicx}
\usepackage{hyperref}
\hypersetup{
colorlinks=true,
linkcolor=red,
citecolor=blue,
filecolor=cyan,
urlcolor=magenta,
}

\journalname{Space Science Reviews}
\newcommand{\aap}{{Astron. Astrophys.}}
\newcommand{\apj}{{Astrophys. J.}}

\newcommand{\solphys}{{Solar Phys.}}
\begin{document}

\title{Oscillator models of the solar cycle}
\subtitle{Towards the development of inversion methods}

\titlerunning{Oscillator models of the solar cycle}        

\author{Il\'idio Lopes
   \and D\'ario Passos
   \and Melinda Nagy
   \and Kristof Petrovay
}


\institute{Il\'idio Lopes$^{1,2}$, D\'ario Passos$^{1,2,3}$, Melinda Nagy$^4$ and  Kristof Petrovay$^4$ \at
              $^1$ Instituto Superior T\'ecnico, Universidade de Lisboa, Portugal \\
              Tel.:  +351 21 8417938,Fax: +351 21 8419118\\
              \email{ilidio.lopes@tecnico.ulisboa.pt; dariopassos@ist.utl.pt}\\
              $^2$ Departmento de  F\'\i sica, Universidade de \'Evora, Portugal\\
              Tel.:  +351 266 745 372, Fax:+351 266 745 394 \\
              $^3$ GRPS, D\'epartment de Physique, Universit\'e de Montr\'eal, \\
              C.P. 6128, Centre-ville, Montr\'eal, Qc, Canada  H3C-3J7\\
              $^4$ Department of Astronomy, E\"otv\"os Lor\'and University, Hungary \\
              Tel.:  +36 1 3722500, Fax:+36 1 3722940 \\
               \email{nagymelinda@caesar.elte.hu; k.petrovay@astro.elte.hu}
}

\date{Received: 15 April 2014 / Accepted: 23 June 2014 / Published: 18 July 2014\\ 
Space Science Reviews \href{http://link.springer.com/article/10.1007/s11214-014-0066-2}{doi:10.1007/s11214-014-0066-2}
}

\maketitle
\begin{abstract}
This article reviews some of the leading results obtained in solar dynamo physics
by using temporal oscillator models as a tool to interpret observational data and dynamo
model predictions. We discuss how solar observational data such as the sunspot number
is used to infer the leading quantities responsible for the solar variability during the last few centuries.
Moreover, we discuss the advantages and difficulties of using inversion methods (or backward methods)
over forward methods to interpret the solar dynamo data.  We argue that this approach could
help us to have a better insight about the leading physical processes  responsible for solar dynamo,
in a similar manner as helioseismology has helped to achieve a better insight on the thermodynamic
structure and flow dynamics in the Sun's interior.
\end{abstract}




\newpage

\section{Introduction}
\label{sec:intro}

The number of dark spots in the Sun's surface has been counted in a systematic way since Rudolf Wolf
introduced the concept,  in the first half of the nineteenth century.
More than any other solar observable,  the sunspot number is considered the strongest signature of the
22-year magnetic cycle. Moreover, since the sunspot number is the longest time series from all
solar observables~\citep{2013Owens},
it makes it the preferred proxy to study the variability and irregularity of the solar magnetic cycle.

In the Sun's interior the large scale magnetic field is generated by a magnetohydrodynamic dynamo
that converts part of the kinetic energy of the plasma motions into magnetic energy. Polarity reversals
occur every 11 years approximately, as it can be observed directly in the Sun's dipolar field, and taking
a full 22-years to complete a magnetic cycle. In fact during each magnetic cycle, the Sun experiences two
periods of maximum magnetic activity, during which magnetic flux tubes created in the tachocline layer,
rise to the Sun's surface by the action of buoyancy, emerging as sunspots pairs \citep{Parker1955}.
The polarity switch is also observed in the change of polarity alignment of these bipolar active regions.

Although we know that the solar dynamo resides within the convection zone, we still don't have a complete
picture where all the physical mechanisms operate \citep{2013JPhCS.440a2014C}. There is a strong consensus
that the physical mechanism behind the production of the large scale toroidal field component, the so called
$\Omega$-effect, is located in the tachocline, a shear layer created by differential rotation and located
at the base of the convection zone.  The major source of uncertainty is the location of the
$\alpha$-effect, the physical mechanism responsible to convert toroidal into poloidal field and close the system.
In truth, this
effect could be in fact a collection of several physical mechanisms that operate at different places and with different
efficiencies. Some examples are the Babcock-Leighton mechanism that operates in the solar surface and
converts the product of decaying active regions into poloidal field, or the action of the turbulent magnetic
helicity that takes place in the bulk of the convection zone. One of the main questions that is still being debated is
the quantification of the importance and relative contribution of each component to the operation of the solar
dynamo. Because different authors choose to give the leading role to one or another $\alpha$ source term, there
is vast number of dynamo models. Most of these are two dimensional models (usually referred as 2.5D because they include
two spatial coordinates plus time) and are constructed using the mean-field theory framework proposed by \cite{Steenbeck1966}.
Despite some short-comes, fruit of the approximations and formulation used, this type of models running in the kinematic regime,
i.e. with prescribed large scale flows, has been very popular within the solar community because they can explain
many of the observable features of the solar cycle.
A detailed discussion on solar dynamo models, stellar magnetism and
corresponding references to the vast literature on this subject can be found in the reviews by
\citet{2010LRSP....7....3C}  and \citet{2009AnRFM..41..317M}.

Another way of tackling the solar dynamo problem is by producing 3D
magnetohydrodynamic (MHD) simulations of the solar convection zone. These
computer intensive simulations solve the full set of the MHD equations (usually
under the anelastic approximation) and are
fully dynamical in every resolved scale, i.e. they take into consideration the
interactions between flow and field and vice-versa -- unlike the kinematic
regime usually used in mean field models, where only the flow influences the
field. Recently these simulations have started to show stable large scale dynamo
behaviour and they are starting to emerge as virtual laboratories for
understanding in detail some of the mechanisms behind the dynamo
\citep{Ghizaru2010, Brown2011, Kapyla2012, PassosCharbonneau2014}.

On the other end of the modelling spectrum, we can find
oscillator  models, that use simplified parameterizations of the main physical
mechanisms that participate in the dynamo process. Although in the Sun's
interior the magnetic field generated by the dynamo has a very rich and complex
structure, as a consequence of the structure of the magnetohydrodynamic
differential equations, some of its main properties can be understood by
analyzing low order differential equations obtained by simplification and
truncation of their original MHD counterparts. Then, several properties of the
cycle that can be extracted by studying these non-linear oscillator models, as
is usually done in nonlinear dynamics. These models have a solid connection to
dynamical systems and are, from the physics point of view the most simple.
This does not mean that they are the easiest to understand because the
reduction in the number of dimensions can sometimes be difficult to interpret
(viz. introduction section of \cite{wilmot-smith2005}). These low order dynamo
models (LODM), as they are some times called, allow for fast computation and
long integration times (thousands of years) when compared to their 2.5D and 3D
counterparts. They can be thought as a first order approximation to study the
impact of certain physical mechanisms in the dynamo solution, or some of the
properties of the dynamo itself as a dynamical system.

The variability exhibited by the sunspot number time series, inspired
researchers to look for chaotic regimes in the equations that describe the dynamo.
For a complete review on this subject consult \cite{Spiegel2009, Weiss2010} and references therein.
Some of the
first applications of LODM were done in this context (e.g. \citet{Ruzmaikin1981, weiss1984, Tobias1995}).
These authors found solutions with cyclic behaviour and variable
amplitude, including extended periods of low amplitude reminiscent of the
grand minima behaviour we see in the Sun. The downside of these initial works
was the fact that although the proposed model equations made sense from a
mathematical point of view, the physics realism they attained was small.
These low order models, with higher or lower degrees of physical
complexity, can be used in many areas and several of their results have been
validated by 2.5D spatially distributed mean field models, which grants them a
 certain degree of robustness. This happens specially in LODM whose formulation
is directly derived from MHD or mean-field theory equations.
Some examples of the results obtained with LODM that have been validated by more
complex mean field models are:
the study of the parameter space, variability and transitions to chaos in
dynamo solutions
\citep{beer1998, wilmot-smith2005, Charbonneau2005a, hiremath2006};
the role of Lorentz force feedback on the meridional flow
\citep{Rempel2006, 2012SoPh..279....1P}; and the influence of stochastic
fluctuations in the meridional circulation and in the $\alpha$-effect
 \citep{CharbonneauDikpati2000, Mininni2001, Mininni2002, LopesPassos2009}.
 Some models even include time delays that embody the spatial segregation
 and communication between the location of source layers of the $\alpha$- and
 $\Omega$-effects. These have been applied to a more general stellar context by
 \cite{wilmot-smith2006} and recently, \cite{Hazra2014, Passos2014a} showed that
 one of this type of time delay LODM that incorporates two different $\alpha$
  source terms working in parallel, can explain how the Sun can enter and exit
  in a self-consistently way from a grand minimum episode.
A couple of LODM even ventured in the "dangerous" field of predictions. For
example \cite{hiremath2008} combined his LODM with an autoregressive model
in order to forecast the amplitude of future solar cycles.

In this article we show how can one of these LODM be used as a
tool to study the properties of the solar magnetic cycle. For this purpose
we use the international sunspot number time series during the past 23
solar magnetic cycles.
Nevertheless, the main focus of this work is to present a strategy inspired
by helioseismology, were an \textit{inversion methodology} is used to infer
variations of some of the LODM parameters over time. Since these parameters
are related to the physical mechanisms that regulate the solar dynamo, this
should in principle, allow for a \textit{first order} reconstruction of the main dynamo
parameters over the last centuries.
In a similar manner to helioseismology, the comparison between model solutions and
data can be done by means of a {\it forward method} in which solar observational data
is directly compared with the theoretical predictions, or by means of a {\it backward method}
in which the data is used to infer the behaviour of leading physical quantities of the theoretical
model. Naturally, it is necessary to develop an inversion technique or methodology
that allows to reconstruct the quantities that have changed during the evolution of
solar dynamo.
This type of studies is well suited to explore several aspects of the solar
and stellar dynamo theory. This can be done by:
\textit{(i)} building a tool to study the dynamo regimes operating in stars;
\textit{(ii)} establishing an inversion methodology to infer the leading quantities responsible for the dynamics and
variability of the solar cycle over time;
\textit{(iii)} comparing the dynamo numerical simulations with the observational data;
\textit{(iv)} use this tool as a toy model to test global properties of the solar dynamo.

Here we particularly focus in discussing the three last items of this list,
with special attention on the development of an inversion method applied here to the sunspot
number time series. This is used to infer some of the dynamics of the solar dynamo back-in-time.
In principle this should allow us to determine the variation profiles of the quantities that
drive the evolution of the magnetic cycle during the last few centuries.

In Section~\ref{sec:LODMLSMF}, we present a non-linear oscillator derived from
the equations of a solar dynamo that is best suited to represent the sunspot number.
In Section~\ref{sec:LODMinversion}, we discuss
how the non-linear oscillator  analogue can be used to invert some of the leading
quantities related with  solar dynamo.
In Section~\ref{sec:SMC} is discussed how solar observational  data
is use to infer properties of the solar magnetic cycle.
In Section~\ref{sec:numericmodels} we present a discussion about how the low order dynamo model
can be used to test the basic properties of modern axisymmetric models and numerical simulations,
as well to infer some leading properties of such  dynamo models.
In Section~\ref{sec:Outlook}, we discuss the outlook for the Sun and other stars.

\section{A LODM for the evolution of the large scale magnetic field}
\label{sec:LODMLSMF}

The basic equations describing the dynamo action in the interior of a star
are obtained from the magnetic-hydrodynamic induction,
and the Navier-Stokes equations augmented by a Lorentz force~\citep{Moffatt:1978tc}.
Under the usual kinematic approximation the dynamo problem consists in finding a flow field with
a velocity $\vec{U}$  that has the necessary properties capable of maintaining the
magnetic field, $\vec{B}$ against Ohmic dissipation ~\citep{2010LRSP....7....3C}.

For a star like the Sun such dynamo models should be able to reproduce well-known
observational features such as: cyclic magnetic polarity reversals with a period of 22 years,
equatorward migration of $\vec{B}$ during the cycle (dynamo wave),
the $\pi/2$ phase lag between poloidal and toroidal components of $\vec{B}$,
the antisymmetric parity across the equator,
predominantly negative/positive magnetic helicity in the Northern/Southern
hemisphere, as well as many of the empirical correlations found in the sunspot records,
like the Waldmeier Rule -- anti-correlation between cycle duration and amplitude;
the Gnevyshev-Ohl Rule -- alternation of higher-than-average and lower-than-average
cycle amplitude and Grand Minima episodes (like the Maunder Minimum) -- epochs
of very low surface magnetic activity that span over several cycles. Given the amount
of complex features that a solar dynamo model has to reproduce, the task at hand
is far from simple.

The vast majority of dynamo models currently proposed to explain the evolution of
the solar magnetic cycle (kinematic mean-field models) became very popular with the
advance of helioseismology inversions and the inclusion of the differential
rotation profile.
In the kinematic regime approximation, the flow field $\vec{U}$ is prescribed and only
the magnetic induction equation is used to determine the evolution of $\vec{B}$.
Generally, the large scale magnetic field, the one responsible for most of the
features observed in the Sun is modelled as the interaction of field and flow where
two source terms ($\Omega$ and $\alpha$) naturally emerge from mean-field theory
~\citep[e.g.,][]{Moffatt:1978tc,Krause:1980vr,2012ApJ...757...71C}.
From the mean-field electrodynamics, the induction equation reads

\begin{equation}
\frac{\partial \vec{B} }{\partial t} =\nabla \times \left(\vec{U}\times \vec{B}
+\alpha \vec{B}-\eta\nabla\times \vec{B}\right),
\label{eq:vecB}
\end{equation}
where $\vec{U}$ is the large-scale mean flow,
and $\eta$ is the total magnetic diffusivity  (including the turbulent diffusivity and the molecular diffusivity). Currently, as inferred from helioseismology,
$\vec{U}$ can be interpreted as a large-scale flow with at least two major flow components,
the differential rotation throughout the solar interior, and the meridional circulation in
the upper layers of the solar convection~\citep{2009LRSP....6....1H}.

Given all the points above, and based on their popularity among the community, we start our
study by considering a reference model based in the kinematic mean-field flux transport framework.
Although the results obtained here are based on this specific type of model, most of the analysis
method used, as well the conclusions reached, can easily be extended to other models.
As usual, under the simplification of axi-symmetry the large-scale magnetic field $\vec{B}$
can be conveniently expressed as the sum of toroidal and poloidal components, that in
spherical polar coordinates $(r,\theta,\phi)$ can be written as

\begin{equation}
\vec{B}(r,\theta,t)=\nabla \times (A_p(r,\theta,t)\vec{e}_\phi) +B_\phi (r,\theta,t) \vec{e}_\phi.
\end{equation}

Similarly, the large-scale flow field  $\vec{U}$ as probed by helioseismology can be
expressed as the sum of an axisymmetric azimuthal (differential rotation) and poloidal
(meridional flow) components:

\begin{equation}
\vec{U}(r,\theta)=\vec{u}_p(r,\theta) + \tilde{r} \Omega (r,\theta) \vec{e}_\phi
\end{equation}
where $\tilde{r}=r\sin{\theta}$, $\Omega$ in the angular velocity
and $\vec{u}_p$ is the velocity of the meridional flow. Accordingly,
such decomposition of $\vec{B}$ (that satisfy the induction equation~\ref{eq:vecB})
and $\vec{U}$ leads to the following set of equations:

\begin{eqnarray}
\frac{\partial A_p}{\partial t}&=&\eta \left(\nabla^2-\tilde{r}^{-2}\right) A_p
-\tilde{r}^{-1}\vec{u}_p\cdot \nabla (\tilde{r}A_p)+\alpha B_\phi
\label{eq:Ap}\\
\frac{\partial B_\phi}{\partial t}&=&\eta \left(\nabla^2-\tilde{r}^{-2}\right)B_\phi
-\tilde{r} \vec{u}_p\cdot \nabla (\tilde{r}^{-1} B_\phi) \nonumber \\
 &+& \tilde{r}\left[\nabla \times (A_p \vec{e}_\phi) \right]\cdot \nabla \Omega
-\Gamma (B_\phi) B_\phi
\label{eq:Bphi}
\end{eqnarray}
where $\eta$ is the magnetic diffusivity and $\alpha$ is the source term of $A_p$
(the mechanism to convert toroidal to poloidal field). Moreover,  following the suggestions of
\citep{Pontieri2003} we also considered that the toroidal field can be removed from the
layers where it is produced by magnetic buoyancy and obeying
$\Gamma\sim\gamma B_\phi^2/8\pi\rho$, where $\gamma$ is a constant related to the removal
rate and $\rho$ is the plasma density.

\subsection{A van der Pol-Duffing oscillator for the solar cycle}

As the Sun's magnetic field changes sign from one solar cycle to the next it is
a plausible idea to attribute alternating signs in odd/even cycles also to other
solar activity indicators such as the sunspot number (SSN). The resulting time
series displays cyclic variations around zero in the manner of an oscillator.
This suggests an oscillator as the simplest mathematical model of the observed
SSN series. As, however, the profile of sunspot cycles is known to be markedly
asymmetric (a steep rise in 3--4 years from minimum to maximum, followed by a
more gradual decline to minimum in $\sim7$ years), a simple linear oscillator
would be clearly a very poor representation of the sunspot cycle. A {\it damped}
linear oscillator
    \begin{equation}
    \label{eq:linosc}\ddot x = -\omega^2x-\mu\dot{x}\\
    \end{equation}
will, on the other hand, naturally result on asymmetric profiles similar to what
is observed. The obvious problem that the oscillation will ultimately decay due
to the damping could be remedied somewhat artificially by applying a periodic
forcing or by reinitializing the model at each minimum. A much more natural way
to counteract the damping, however, is the introduction of nonlinearities into
the equation ---indeed, such nonlinearities are naturally expected to be present
in any physical system, see below. As long as the nonlinearity is relatively
weak, the parameters $\omega^2$ and $\mu$ can be expanded into Taylor series
according to $x$. Due to the requirement of symmetry (i.e. the behaviour of the
oscillators should be invariant to a sign change in $x$) only terms of even
degree will arise in the Taylor series. To leading order, then, we can
substitute
    \begin{equation}
    \omega^2 \rightarrow \omega^2-\lambda x^2 \qquad
    \mu \rightarrow \mu (\xi x^2 - 1)  \\
    \end{equation}
into equation (\ref{eq:linosc}) resulting in
    \begin{equation}
    \label{eq:nonlinosc}\ddot x = -\omega^2x-\mu(\xi x^2 - 1)\dot{x}
    +\lambda x^3\\
    \end{equation}

In the particular case when $\lambda=0$ (i.e. the nonlinearity affects the
damping only) and the other parameters are positive, the system described by
equation (\ref{eq:nonlinosc}) is known as a van der Pol oscillator. The alternative
case when nonlinearity affects the directional force/frequency only, i.e.
$\xi=0$, $\mu<0$ and $\lambda\neq 0$, in turn, represents a {\it Duffing
oscillator}. Due to their simplicity and universal nature these two systems are
among those most extensively studied in nonlinear dynamics. It is
straightforward to see that the oscillator is non-decaying, i.e. the origin is
repeller, whenever $\mu>0$ (negative damping) in the case of a van der Pol
oscillator and/or $\omega^2>0$ and $\lambda>0$ in a Duffing oscillator.When a
nonlinearity is present in both paramters (i.e. $\lambda$ and $\xi$ are both
non-zero) a combined {\it van der Pol-Duffing oscillator} results.

The van der Pol--Duffing oscillator, however, is more than just a good heuristic
model of the solar cycle. In fact, an oscillator equation of this general form
can be derived by a truncation of the dynamo equations. As noted before, we are
especially interested in capturing the temporal dynamics associated with the
large scale magnetic field. In order to construct a low order model aimed at
capturing this dynamics, we follow the procedures described in
\citet{PassosLopesApJ2008, PassosLopesJASTP2011}.

It has been suggested by \citet{Mininni2000, Mininni2001} 
and \citet{Pontieri2003}  that a
dimensional truncation of the dynamo equations (\ref{eq:Ap}) and (\ref{eq:Bphi})
is an effective method to reduce the system's dimensions and capture phenomena
just on that scale. Following that ansatz, gradient and laplacian operators are
approximated by a typical length scale of the system $l_0$ (e.g. convection zone
length or width of the tachocline), leading to $\nabla \sim l_o^{-1}$ and
$\nabla^2\sim l_o^{-2}$. Analogously this can be interpreted as a collapse of
all spatial dimensions, leaving only the temporal behaviour. In terms of
dynamical systems, we are projecting a higher dimensional space into a single
temporal plane. After grouping terms in $B_\phi$ and $A_p$ (now functions only
dependent of the time) we get
\begin{eqnarray}
    \frac{d B_\phi}{d t} &=& c_{1} B_\phi + c_{2} A_p - c_{3} B_\phi^3
    \label{eq:dBdt} \\
    \frac{d A_p}{d t} &=& c_{1} A_p + \alpha \, B_\phi
    \label{eq:dAdt}
\end{eqnarray}
where we have defined the \textit{structural coefficients}, $c_n$, as
\begin{eqnarray}
    c_{1}&=&\eta \left(\frac{1}{l_0^2} -
    \frac{1}{\bar{r}^2} \right) - \frac{v_p}{l_0}
    \label{eq:c1} \\
    c_{2}&=& \frac{\bar{r} \Omega}{l_0^2}
    \label{eq:c2} \\
    c_{3} &=& \frac{\gamma}{8 \pi \rho}
    \label{eq:c3}
\end{eqnarray}

We now concentrate in creating an expression for the time evolution of $B_\phi$ since it is
the field component directly associated with the productions of sunspots. We derive
expression (\ref{eq:dBdt}) in order to the time, and substitute (\ref{eq:dAdt}) in it to
take away the $A_p$ dependence yielding

\begin{equation}
    \frac{d^2 B_\phi }{d t^2} + \omega^2 B_\phi +
    \mu (3 \xi B_\phi^2 - 1)\frac{\partial B_\phi}{\partial t} - \lambda B_\phi^3 = 0,
    \label{eq:vdp}
\end{equation}
where $\omega^2=c_1^2 - c_2 \alpha$, $\mu = 2c_1$, $\xi=c_3 / 2 c_1$ and
$\lambda=c_1 c_3$ are model parameters that depend directly on the structural
coefficients. The name used to describe $c_n$ comes from the fact that these
coefficients contain all the background physical structure (rotation, meridional
circulation, diffusivity, etc.) in which the magnetic field evolves.

\begin{figure}
\includegraphics[scale=0.7]{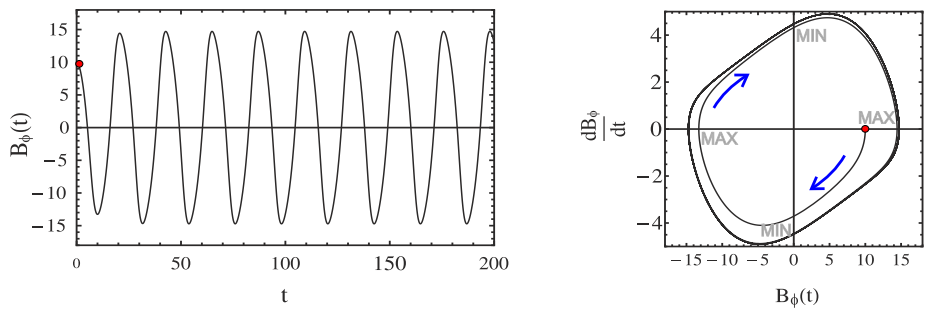}
\caption{Left panel represents the time evolution of $B_\phi(t)$ obtained from equation
(\ref{eq:vdp}) with parameters  $c_1=0.08$, $c_2=-0.09$, $c_3=0.001$, $\alpha_0=1$.
In the right we have a ($B_\phi$, $dB_\phi/dt$)
phase space representation of the solution. The blue arrows indicate the direction of
increasing time and the red dot the initial value used. Also indicated in this panel are
the regions corresponding to the maxima and minima of the cycle. \textit{Adapted from}
\citet{PassosLopesJASTP2011}.}
\label{fig:1}
\end{figure}

This oscillator (equation~\ref{eq:vdp}) is a van der Pol-Duffing oscillator and
it appears associated with many types of physical phenomena that imply
auto-regulated systems. This equation is a quite general result which
$\vec{B}$ should satisfy. In this case, unlike in the classical van der
Pol-Duffing oscillator, the parameters are interconnected by a set of
relations that link the present oscillation model with the original set of
dynamo equations (\ref{eq:Ap}-\ref{eq:Bphi}). This interdependency between
parameters will eventually constrain the solution's space. As in the classical
case, $\omega$ controls the frequency of the oscillations or the period of the
solar magnetic cycle, $\mu$ controls the asymmetry (or non-linearity) between
the rising and falling parts of the cycle and $\xi$ affects directly the
amplitude. The $\lambda$ parameter, related to buoyancy loss mechanism
sets the overall amplitude peak amplitude of the solution. Figure (\ref{fig:1})
shows the solution of equation (\ref{eq:vdp}) in a time vs. amplitude diagram
(left) and in a $\{B_\phi$, $dB_\phi/dt\}$ phase space. From this figure we find
that this dynamo solution, under suitable parametrization (viz. next section) is
a self-regulated system that rapidly relaxes to a stable 22-year oscillation. In
the phase space the solution tend to a limit cycle or attractor. A complete
(clockwise) turn in the phase space corresponds to a complete solar magnetic
cycle.

\subsection{A semi-classical analysis method using a non-linear oscillator}

In order to estimate values for the coefficients in equation (\ref{eq:vdp}),
\citet{Mininni2000} and \citet{PassosLopesApJ2008} fitted this oscillator model
either to a long period of the solar activity (several solar cycles) or to each
magnetic cycle individually. We shall return to this point in subsequent
sections. A more general approach to the problem of finding the parameter
combinations with which the classical van der Pol-Duffing oscillator returns
solar-like solutions was taken by \citet{nmpk2013}. The authors mapped the
parameter space of the oscillator by adding stochastic noise to its
parameters using different methods. The objective was to constrain the
parameter regime where this nonlinear model shows the observed attributes of the
sunspot cycle, the most important requirement being the presence of the
Waldmeier effect \citep{Waldmeier} according to the definition of
\citet{cameron}.

Noise was introduced either as an Ornstein--Uhlenbeck process \citep{Gillespie96}
or as a piecewise constant function keeping a constant value for the interval of
the correlation time. The effect of this noise was assumed to be either additive
or multiplicative. The amplitudes and correlation times of the noise defined
the phase space. The attributes of the oscillator model were
first examined in the case of the van der Pol oscillator (no Duffing cubic term)
with perturbation either in the nonlinearity parameter,  $\xi$ or the damping
parameter, $\mu$, as shown in the equations below:

    \begin{eqnarray}
    \label{e:vdp-mu}\ddot x&=&-\omega_0^2x-\mu(t)\left[ \xi_0 x^2-1\right]\dot{x}\\
    \label{e:vdp-xi}\ddot x&=&-\omega_0^2x-\mu_0\left[ \xi(t) x^2-1\right]\dot{x}.
    \end{eqnarray}

\begin{figure}[htb!]
\center
\includegraphics[width=\textwidth]{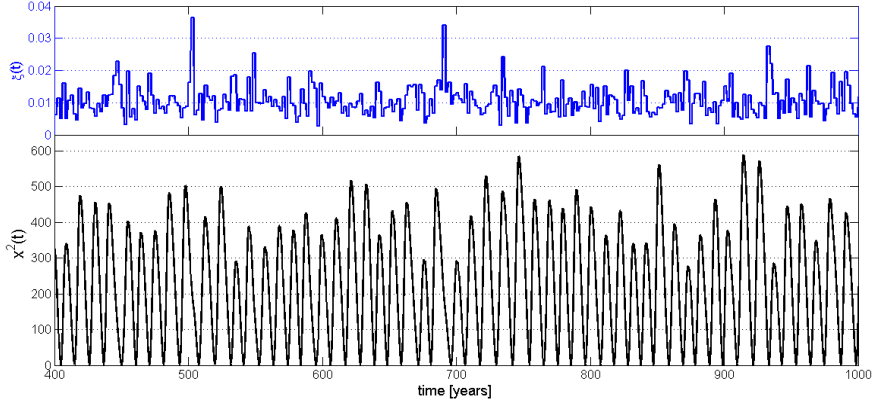}
\caption{Bottom panel: a 600 year long time series resulting from a
stochastically perturbed van der Pol oscillator stochastically perturbed in one
of its parameters (nonlinearity $\xi(t)$). SSN values were defined as
$x^{2}(t)$. The noise applied (piecewise constant in this case) is shown in the
top panel.}\label{fig:nmpk2013}
\end{figure}

The constant parameters, $\omega_0$, $\mu_0$ and $\xi_0$ used were taken from
the fitted values listed by \citet{Mininni2000}. Note that in this simple case,
variations in these parameters were assumed to be independent from each other,
whereas in reality they are interrelated (see equation~\ref{eq:vdp}). The
results show that the model presents solar-like solutions when a multiplicative
noise is applied to the nonlinearity parameter,
as in equation (\ref{e:vdp-xi}). An example of a time series produced by this
type of oscillator is shown in Figure (\ref{fig:nmpk2013}).

\begin{figure}[htb!]
\center
\includegraphics[width=\textwidth]{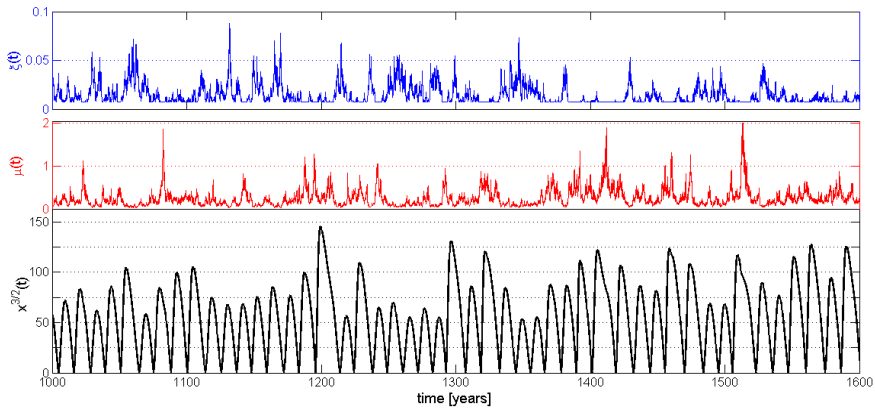}
\caption{Bottom panel: a 600 year long time series resulting from a van der
Pol--Duffing oscillator stochastically perturbed in two of its parameters, $\xi$
and $\mu$. SSN values were here defined as $x^{3/2}(t)$, following
\citet{bracewell1988}. The noise applied is shown in the top and middle panels.
The Duffing parameter was here given a constant value
$\lambda_0=5\times10^{-5}$.}\label{fig:nmpk201x}
\end{figure}

As a next step towards a fully general study, let us consider the case where
both $\xi$ and $\mu$ are simultaneously perturbed and the Duffing term $-\lambda
x^3$ is also kept in the oscillator equation (\ref{eq:nonlinosc}).  Noise is
applied to $\mu$ but it also affects $\xi$ as the values of $\xi$ and $\mu$ are
assumed to be related as
\begin{eqnarray}
\xi(t)=\frac{C_{\xi}}{\mu(t)} \;\;\; \mathrm{and} \;\;\; \lambda(t)=C_{\lambda}\frac{\mu(t)}{2},
\end{eqnarray}
\citep{PassosLopesApJ2008}; here, $C_{\xi}$ and $C_{\lambda}$ are constants.

A mapping of the parameter space shows that in this case solar-like solutions
are more readily reproduced compared to the case when only one parameter was
assumed to vary (see Fig.~\ref{fig:nmpk201x}). This finding is in line with the information derived from the
LODM developed in the previous section. An ongoing study shows that the
corresponding time dependence in the Duffing parameter, as predicted by the
LODM, has a significant effect on the character of the solution.

We note that additive noise was first applied to one parameter ($\xi$) of a van
der Pol oscillator by \citet{Mininni2000, Mininni2001} but the focus of that work
was on reproducing cycle to cycle fluctuations, withour considering the
Waldmeier effect. This model was further analysed by \citet{Pontieri2003} (and
references therein) who studied the behavior of the Hurst exponent of this
system and concluded that this type of fluctuations implies that the stochastic
process which underlies the solar cycle is not simply Brownian. This means that
long-range time correlations could probably exist, opening the way to the
possibility of forecasts on time scales comparable to the cycle period. An
attempt to introduce the effects of such non-gaussian noise statistics into the
LODM was made by \citet{Vecchio+Carbone} who suggest that this may contribute to
cyclic variations of solar activity on time scales shorter than 11 years.

\section{Coupling a LODM with observational data - inversions}
\label{sec:LODMinversion}
\begin{figure}[htb!]
\centering
\includegraphics[width=\textwidth]{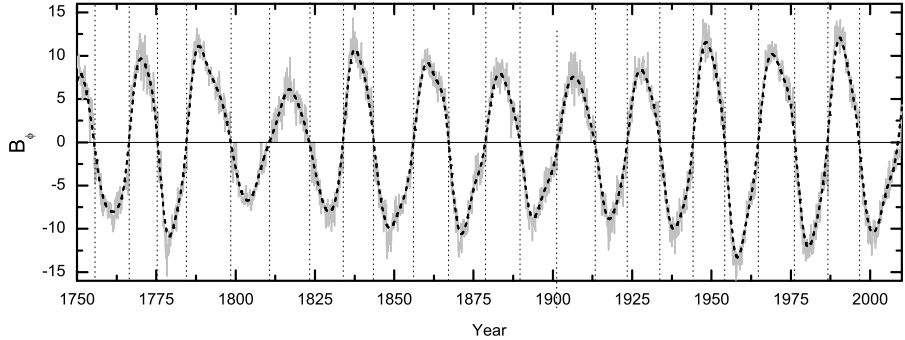}
\caption{Black dashed line represents the built proxy for the toroidal field. $B_\phi$
is obtained by calculating $\sqrt{SSN}$, changing the sign of alternate cycles (represented in
gray), and smoothing it down using an FFT low pass filter of 6 months. The vertical thin
dotted  lines represent solar cycle minima.}
\label{fig:2}
\end{figure}

In the previous example a perturbation method was studied in order to find
solar like solutions for this non-linear oscillator. Another way of thinking is to
pair the oscillator directly to some solar observable and try to constraint its
parameters. As mentioned in the introduction we choose the international sunspot number,
$SSN$ and we use it to build a proxy of the toroidal magnetic component.
Since the $SSN$ is usually taken to be proportional to the toroidal field
magnetic energy that erupts at the solar surface ($\propto B_\phi^2$),
\citet{1995MNRAS.273.1150T}, this makes it ideal for compare with solutions of the LODM.
Taking this in consideration, \citet{Mininni2000, PassosLopesApJ2008} have built
a toroidal field proxy based on the sunspot number by following the procedure proposed
 by~\citet{1996ApL&C..34...35P}, i.e. $B_\phi\approx\pm\sqrt{SSN}$. Details about
 the construction of the toroidal proxy (see Fig.~\ref{fig:2}) can be found in
 \citet{Mininni2001, PassosLopesApJ2008} and \citet{Passos2012}.

\subsection{The averaged behaviour of the solar dynamo attractor}

The solution obtained for equation (\ref{eq:vdp}) presented in Figure (\ref{fig:1}) shows that the
solar cycle is a self-regulated system that tends to a stable solution defined by an attractor
(limit cycle). If we allow for the different physical processes responsible
for the solar dynamo and embedded in the structural coefficients (\ref{eq:c1}, \ref{eq:c2} and
\ref{eq:c3}), i.e. the differential rotation, the meridional circulation flow, the $\alpha$ mechanism,
and the magnetic diffusion, to change slowly from cycle to cycle then we start to observe deviations
 from the equilibrium state. If deviations from this sort of dynamical balance occur, such that if
 one of these processes changes due to an external cause, the other mechanisms also change to
 compensate this variation and ensure that the solar cycle finds a new equilibrium.

\begin{figure}
\centering
\includegraphics[scale=0.50]{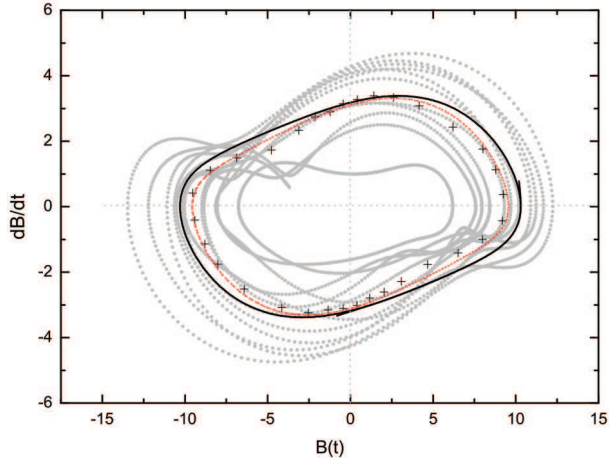}
\caption{Phase-space diagram for the toroidal proxy $B_\phi(t)$ :
The crosses correspond to local area averaged values found by dividing the
data into 32 temporal intervals. The red dashed curve is a fit to the crosses.
The continuous black curves correspond to a fit to all data points (not grouped in
intervals). For the red curve we have that $\mu=0.1645$, $\omega=0.3523$,
$\xi=0.0147$ and $\lambda =0.0005$. Figure \textit{adapted from}~\citet{2008SoPh..250..403P}.}
\label{fig:4}
\end{figure}

To test this idea of an equilibrium limit cycle, we fit the LODM parameters to the
\{$B_\phi$,$dB_\phi/dt$\} phase space of the built toroidal proxy (see Fig.~\ref{fig:4}).

If one of the parameter's variation is very large the system can be dramatically affected,
leading to a quite distinct evolution path like the ones found during the solar grand
minima. We will develop this subject in a subsequent section.

\subsection{Matching of solutions to observable characteristics of the solar cycle}

Solutions with fluctuation similar to those we see on the solar cycle are easily set by
variations in the $\mu$  parameter (and the physical processes associated with it). By
definition the structural coefficient that regulates this parameter ($c_1$) also has
an important role in the other parameters ($\omega$, $\xi$ and $\lambda$).
In the LODM equation (\ref{eq:vdp}) the $\mu$
and $\mu\xi$ quantities regulate the strength and the non-linearity of the damping.
Moreover, an occasional variation on $\mu$, like a perturbation on the meridional
flow amplitude, $v_p$  (see structural coefficients~\ref{eq:c1}) will affect all
sets of parameters
leading to the solar dynamo (equation~\ref{eq:vdp}) to find a new equilibrium,
which will translate into the solar magnetic cycle observable like the sunspots number,
showing an irregular behaviour.

The well-known relation discover by Max Waldmeier~\citep{1935MiZur..14..105W},  that  the time
that the sunspot number takes to rise from minimum to maximum is inversely proportional  to the
cycle amplitude in naturally captured by the LODM assuming discrete variations in $\mu$.
Notice that the Waldmeier effect occurs as a consequence of the limit cycle becoming increasingly sharp
as $\mu$ increases, i.e., the sunspot number amplitude increases as the cycle's rising times gets
shorter.

\section{How to infer properties of the Solar Magnetic Cycle}
\label{sec:SMC}

From the physical point of view, based on observations, we know that in the Sun some of the physical background structures that are taken as constant in our standard dynamo solution aren't so. In order to test that specific changes of the background state lead to the observed changes in the amplitude of the solar cycle, the following strategy was devised. At a first approximation we assume that the structural coefficients can change only discretely in time, more specifically from cycle to cycle while the magnetic field is allowed to evolve continuously. The idea is that changing coefficients will generate theoretical solutions with different amplitudes, periods and eigen-shapes at different times and by comparing these different solution pieces with the observed variations in the solar magnetic field, we are able to infer information about the physical mechanisms associated with the coefficients.

\begin{figure}[htb!]
    \centering
    \includegraphics[width=\textwidth]{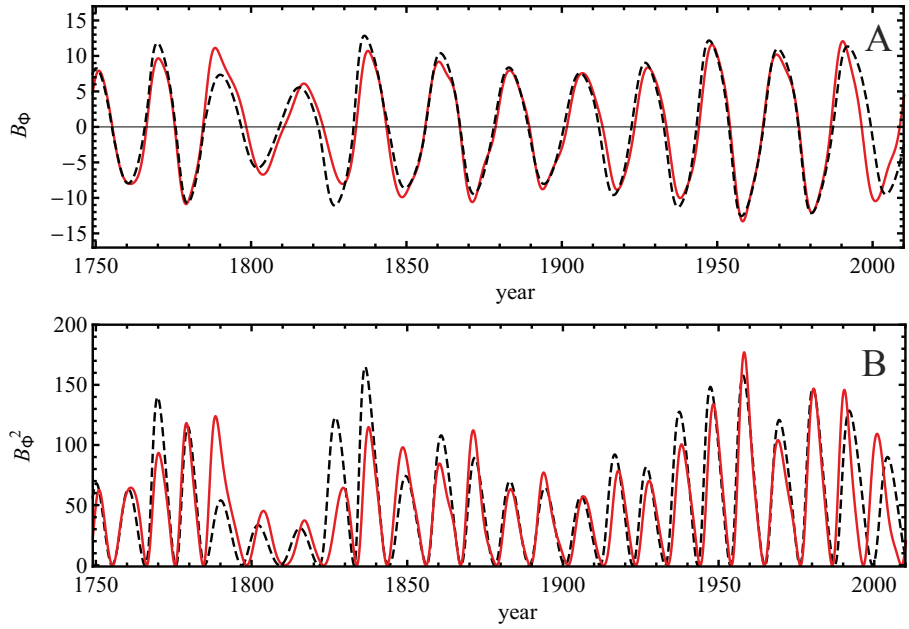}
    \caption{\small \textbf{A)} theoretical solution (red) obtained by fitting it the proxy at each individual cycle. \textbf{B)} Direct comparison of the model behavior (red) and observed solar cycle amplitude (black). In panel B) we just plot the squared of panel B) and this also amplifies the differences between both curve. \textit{Adapted from} \citet{Passos2012}.}
    \label{fig:fit}
\end{figure}

To do this we compare our theoretical solution with a proxy built from the International monthly averaged SSN since 1750 to the present. As mentioned before we assume that $B_\phi \propto \sqrt{SSN}$.
The proxy data is separated into individual cycles and fitted using equation (\ref{eq:vdp}), considering that the buoyancy properties of the system are immutable, i.e. $c_3$ is constant throughout the time series. This means that when we fit the LODM to solar cycle $N$, we will retrieve the set of $c_{nN}$ coefficients that best describe that cycle. This allows to probe how these coefficients vary from cycle to cycle and consequently how the physical mechanisms associated with them evolve in time \citep{LopesPassos2009, Passos2012}.
Equation (\ref{eq:vdp}) is afterwards solved by changing the parameters to their fit value, at every solar minimum using a stepwise function (similar to that presented in
figure (\ref{fig:vp_c1_var}) for $c_1$). Figures (\ref{fig:vp_c1_var}) and (\ref{fig:fit}) highlight this procedure.
The fact that such a simplified dynamo model can get this degree of resemblance with the observed data just by controlling one or two parameters is an indication that it captures the most important physical processes occurring in the Sun.

\bigskip

\subsection{Meridional circulation reconstruction}

The simple procedure previously described allows to reconstruct the behavior of solar parameters back in time.
Using an improved fitting methodology, \citet{Passos2012} obtained with this model the reconstruction of the
variation levels of the solar meridional circulation for every solar (sunspot) cycle over the last 250 years.
One must notice that in this specific LODM the amplitude of the cycle depends directly on the amplitude
of the meridional flow during the previous cycle. It is completely possible to imagine that other models that
consider a different theoretical setup might return a different behaviour.

\begin{figure}[htb!]
    \centering
    \includegraphics[width=\textwidth]{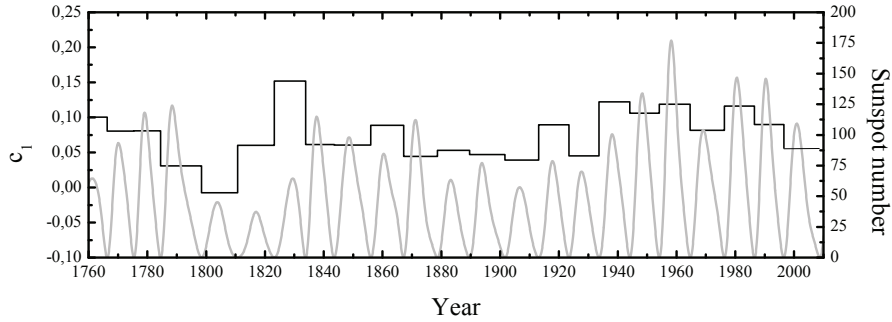}
    \caption{\small Reconstruction of the meridional circulation profile represented by $c_1$ (black line) compared to smoothed the sunspot number (gray).  \textit{Adapted from} \citet{Passos2012}}
    \label{fig:vp_c1_var}
\end{figure}

Looking at equation (\ref{eq:c1}), we can see that the coefficient $c_1$
 depends on two physical parameters, the magnetic diffusivity, $\eta$,
and the amplitude of the meridional circulation, $v_p$. The magnetic
diffusivity of the system is a property tightly connected with turbulent convection and is generally believed to change only in time scales of the order of stellar evolution. This leaves variations in $v_p$ as the only
plausible explanation for the variation observed from cycle to cycle.
Therefore, by looking at the evolution of $c_1$ we can can effectively assume that
we are looking at the variation in the strength of meridional circulation.
The results obtained are presented in Figure (\ref{fig:vp_c1_var}).

\bigskip

\begin{figure}[htb!]
    \centering
    \includegraphics[width=\textwidth]{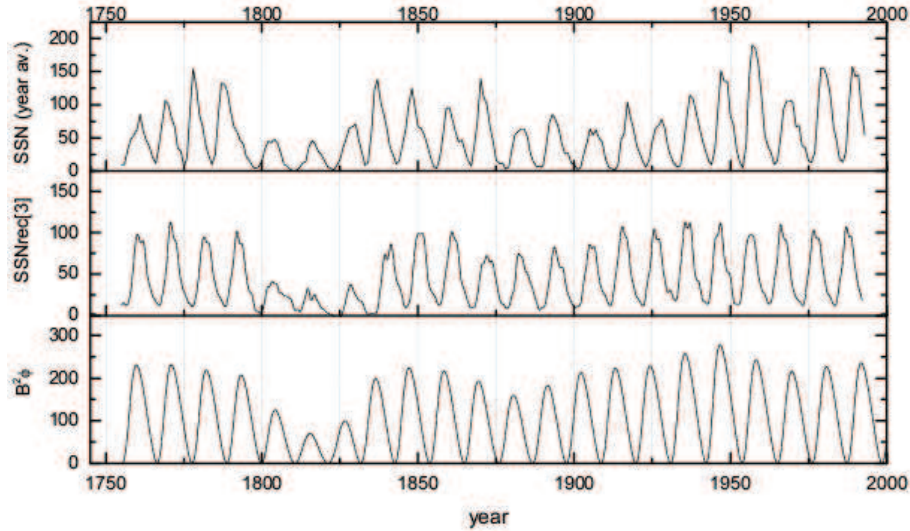}
    \caption{\small Comparison between the observed solar cycle amplitude (top), a sunspot reconstruction using the \texttt{Surya Dynamo Code} (middle) and the LODM results (bottom). In these simulations were only considered variations in the meridional flow amplitude every 2 sunspot cycles (1 magnetic cycle). \textit{Adapted from} \citet{PassosLopesApJ2008}, \citet{LopesPassos2009} and \citet{PassosTese}.}
    \label{fig:LODM_Surya_comp}
\end{figure}

Although this result is in itself interesting, a more important
concept came from this study. When \citet{PassosLopesApJ2008} presented
their results for the first time, they introduced the idea that
coherent long term variations (of the order of the cycle period) in
the strength of the meridional circulation could provide an
explanation for the variability observed in the solar cycle (see Fig.~\ref{fig:LODM_Surya_comp}). This
result was also \textit{a posteriori} numerically validated using a
2.5D flux transport model (\citet{LopesPassos2009} and \citet{Karak2010}).
Only a couple of years later, \citet{HathawayScience2010} presented meridional circulation
measurements spanning over the last solar cycle. Their measurements
confirmed that the amplitude of this plasma flow changes considerably
from cycle to cycle.

Recently two other groups have tested this idea with their 2.5D dynamo
models finding additional features based in this effect,
c.f.~\citet{Karak2010}, \citet{Karak2011} and \citet{NandyNature2011}. For example it was
found that the instant at which the change in the meridional flow
takes place, has an influence in the duration of the following solar
cycle. This was used as an explanation for the abnormally long
duration of the last minimum. Just for reference, the numbering of
solar cycles only started after 1750 with solar cycle 1 beginning in
1755. At this moment we are in the rising phase of solar cycle 24.

\subsection{Explaining solar grand minima}

\subsubsection{Variations in the meridional flow}

\begin{figure}[htb!]
    \centering
    \includegraphics[width=\textwidth]{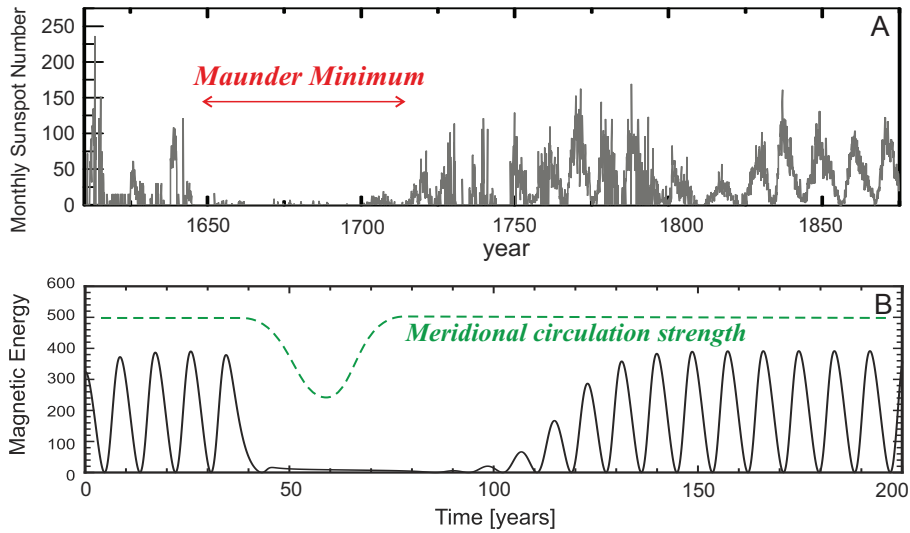}
    \caption{\small Top panel represents the sunspot measurements
where the Maunder Minimum period is highlighted by the red arrow. In
the bottom we show the response of the LODM (magnetic energy in
arbitrary units) to a decrease in the strength of the meridional flow
exemplified by the green dashed line.  \textit{Adapted from}
\citet{PassosLopesJASTP2011}}
    \label{fig:grandmin}
\end{figure}

\begin{figure}
\includegraphics[scale=0.50]{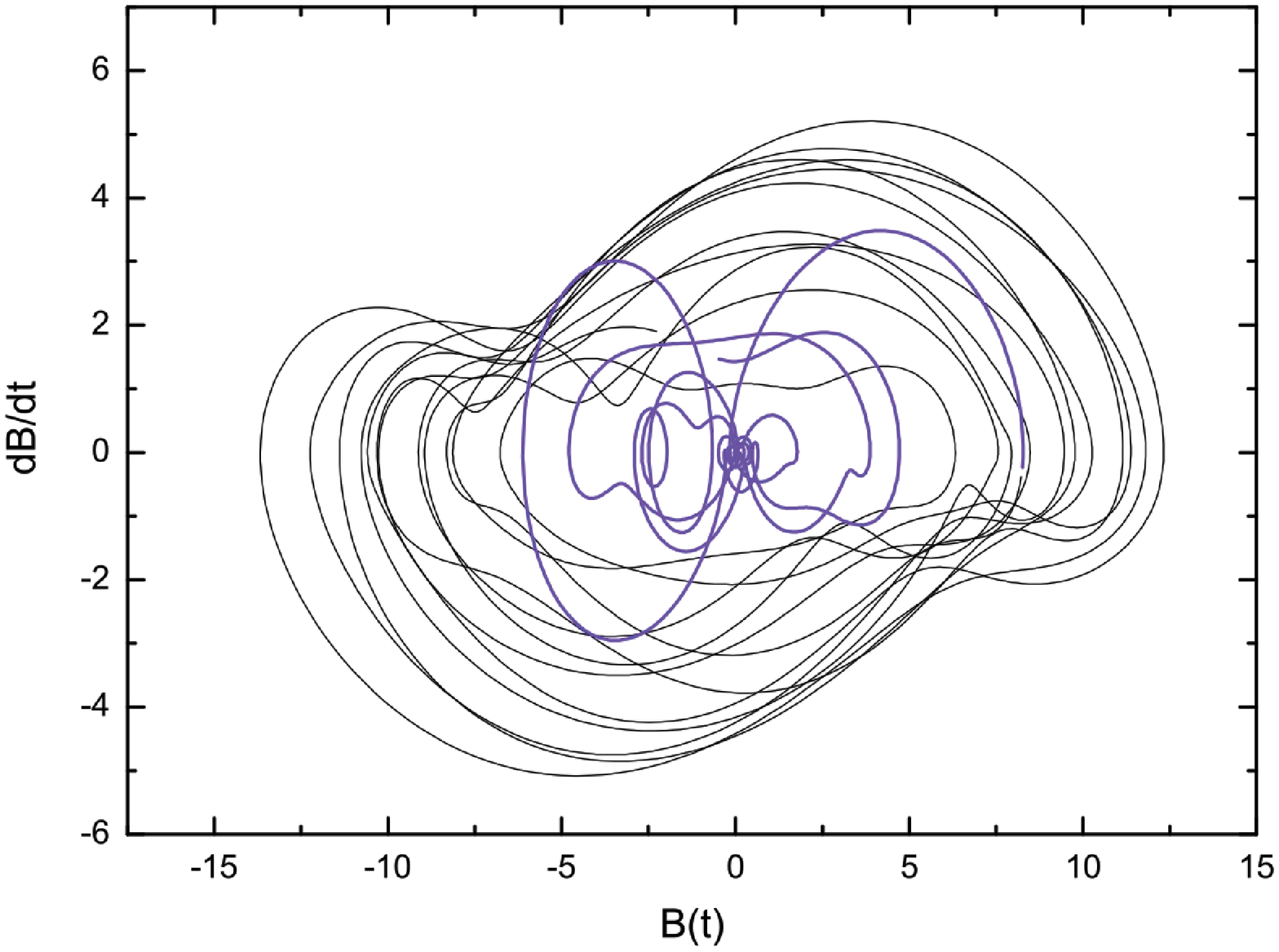}
\caption{The phase-space diagram for B(t) corresponds to the value obtained
from the sunspot number temporal series as in Figure (\ref{fig:grandmin}).
\textit{Adapted from} \citet{PassosLopesJASTP2011} and \citet{PassosTese}. }
\label{fig:Bgrandmin}
\end{figure}

Solar grand minima correspond to extended periods (a few decades) where very low or no solar activity occurs. During theses periods no sunspots (or very few) are observed in the solar photosphere and it is believed that other solar phenomena also exhibit low levels of activity. The most famous grand minimum that has been registered is the Maunder Minimum which occurred between the years of 1645 and 1715 (\cite{Eddy1976}).

A possible explanation for the origin of these quiescent episodes was
put forward by \citet{PassosLopesJASTP2011}. Using a LODM, they
showed that a steep decrease in the meridional flow amplitude can lead
to grand minima episodes like the Maunder minimum (see Figs.~\ref{fig:grandmin} 
and \ref{fig:Bgrandmin}). This effect presents the
same visual characteristics as the observed data, namely a rapid
decrease of magnetic intensity and a gradual recovery into normal
activity (see Fig.~\ref{fig:grandmin}) after the meridional
circulation amplitude returns to its normal values. A similar result
was later obtained by \citet{Karak2010}, again using a more complex
2.5D numerical flux transport model. Nevertheless the reasons that
could lead to a decrease of the meridional flow amplitude were not
explored. This served as a motivation to study the behavior of this
LODM in the non-kinematic regime, explained in section \S 5.2.

\bigskip

\subsubsection{Fluctuations in the $\alpha$ effect}

Some examples mentioned in the introduction, hint that fluctuations in the $\alpha$ mechanism
can also trigger grand minima. We focus on a specific example now, the LODM developed by
\citet{Hazra2014}. In this work the authors used a time-delay LODM
similar to that presented in \citet{wilmot-smith2006} but expanded with the addition of a second
$\alpha$ effect. This model incorporates two of these mechanisms, one that mimics the surface
Babcock-Leighton mechanism (BL), and another one analogous to the classical mean-field (MF)
$\alpha$-effect that operates in the bulk of the convection zone. This set up captures the
idea that the BL mechanism should only act on strong magnetic fields that reach the surface,
and that weak magnetic fields that diffuse through the convection zone should feel the
influence of the MF $\alpha$. The authors subject these two effects to different levels
of fluctuations and find that in certain parameter regimes, the solution of the system shows
the same characteristics as a grand minimum. These results were also validated by implementing
a similar set up into a 2.5D mean-field flux transport dynamo model \cite{Passos2014a}. Again
this shows the usefulness of low order models to probe ideas before their implementation into
more complex models.

\subsection{Solar cycle predictability}

\begin{figure}[htb!]
    \centering
    \includegraphics[width=\textwidth]{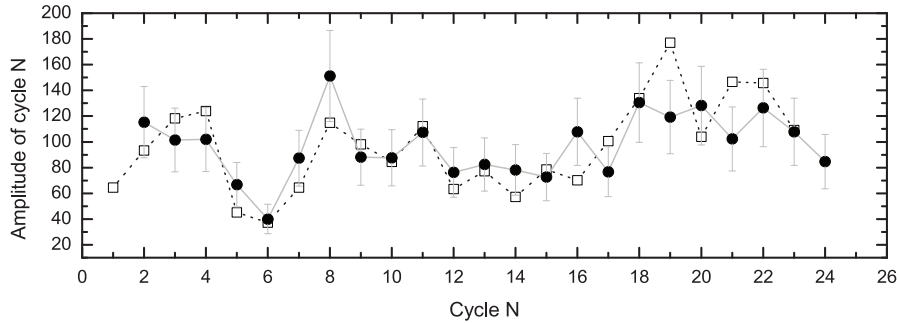}
    \caption{Observed values (white squares) and predicted values black circles with gray error bars for
    the cycle amplitude. The red circle is the predicted amplitude for solar cycle 24 based on this
    methodology. \textit{Adapted from} \citet{Passos2012}}.
    \label{fig:predict24}
\end{figure}

For the near future, perhaps one of most interesting applications of this LODM is its use in the predictability of future solar cycles amplitudes. The first step towards this objective is presented in~\citet{Passos2012}.
The authors studied the correlations between the LODM fitted structural coefficients and cycle's characteristics (amplitude, period and rising time). They found very useful relationships between these quantities measured for cycle N and the amplitude of cycle N+1. These relationships were put to the test by predicting the amplitude of current solar cycle 24 (see Fig.~\ref{fig:predict24}).

\section{LODM as probe of numerical dynamo models}
\label{sec:numericmodels}

In recent years there have been strong developments of different types of dynamo models
to compute the evolution of the solar magnetic activity and to explore some of the causes of magnetic variability.
Two classes of models have been quite successful, the kinematic
dynamo models
and, more recently, the global magnetohydrodynamical models.
Both types of dynamo models have a quite distinct approach to the dynamo theory,
the first one resolves the induction magnetic equation for a prescribed velocity field
(which is consistent with helioseismology), and the second one obtains global magnetohydrodynamical
simulations of the solar convection zone.
Many of these models are able to reproduce some of the many observational features
of the solar magnetic cycle.  Nevertheless,  it remains quite a difficult task to successfully
identify  which are the leading physical processes in current dynamo models
that actually drive the dynamo in the solar interior.
The usual method to test these dynamo models is to compare their theoretical predictions
with the different sets  of data, including the sunspot numbers, however, in many cases
the conclusions obtained are very limited,  as different physical mechanisms
lead to very identical predictions. This problem also arises in the comparison
between different dynamo models, including different types of numerical simulations.

A possible solution to this problem is to use inverted quantities (obtained form observational data)
to test the quality of the different solar dynamo model, rather than making direct comparison of data.
For those of you familiarized with helioseismology,  there is a good analogue:
it is the equivalent to compare the inverted sound speed profile (obtained from observational data)
with the sound speed profile predicted by solar models ({\it backward approach}),
rather than compare predicted frequencies with observational frequencies ({\it forward approach}).
The former method to test physical models is more insightful than the latter one. At the present
level of our understanding  of the solar dynamo theory, as a community we could gain a more profound
understanding of the mechanisms behind  the solar magnetic variability, if we
start developing some backward methods to analyse solar observational data and test dynamo models.

\begin{figure}
\includegraphics[scale=0.50]{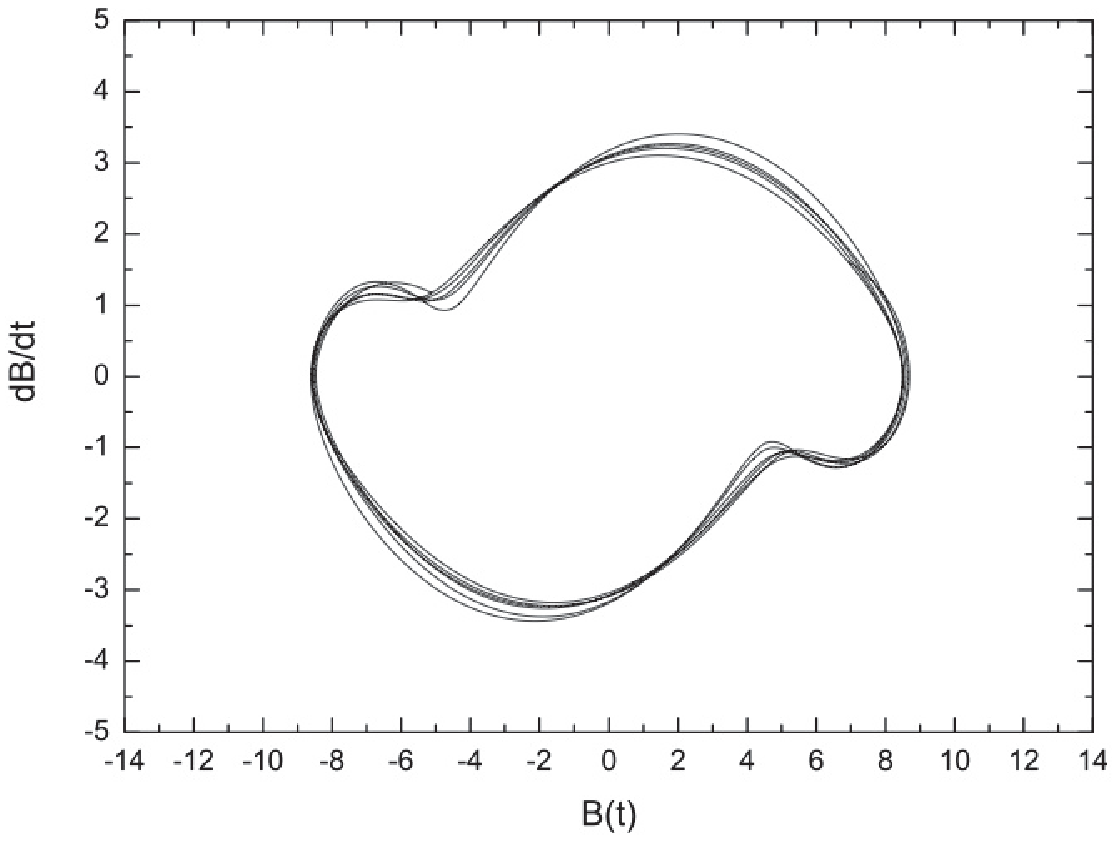}\includegraphics[scale=0.48]{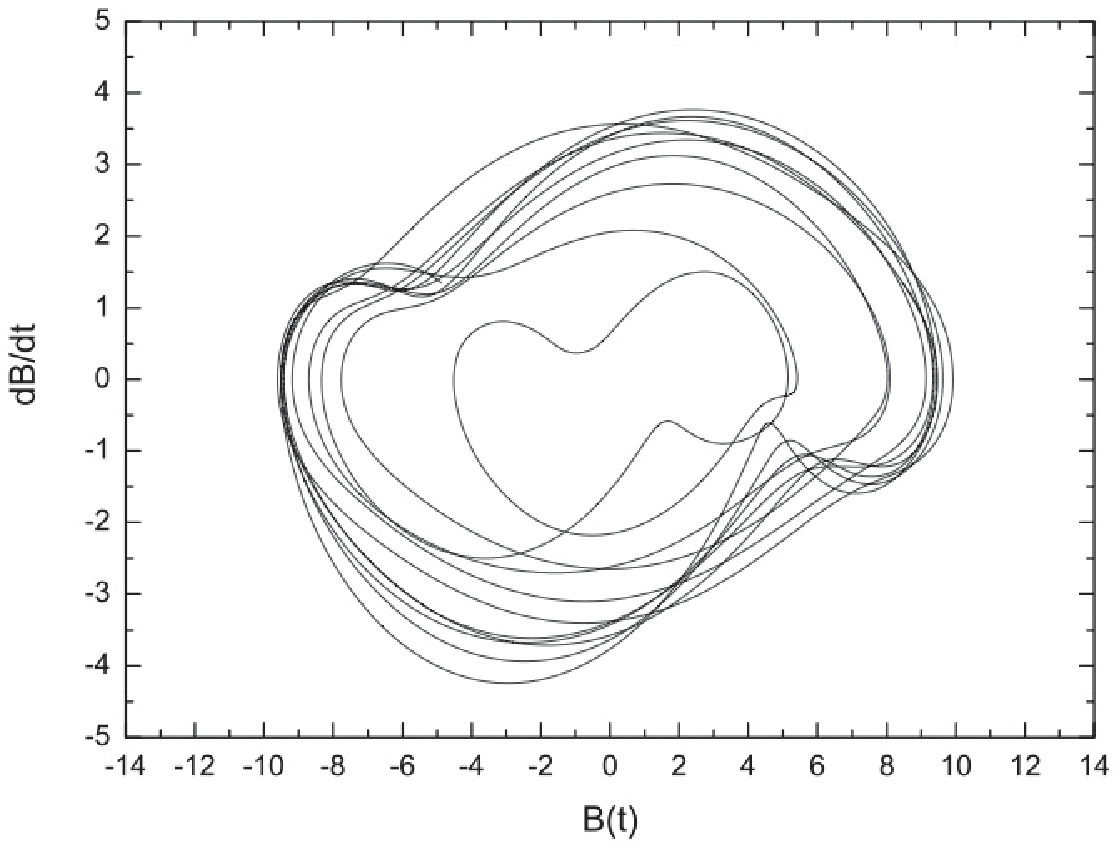}
\caption{The phase-space diagram for B(t) of a kinematic dynamo model
(using the Surya's kinematic model)): (a) a standard kinematic dynamo model ($v_p$ is constant);
(b) A variable kinematic dynamo model in which the  $v_p$ for each magnetic cycle
corresponds to the value obtained from the sunspot number temporal series.
Both simulations correspond to 130-year time series. The small variability present in the
left panel is due to the stabilization of the numerical solution.
Figure \textit{adapted from}~\citet{2009SoPh..257....1L}.
}
\label{fig:compsuryssn}
\end{figure}

\subsection{Using a LODM in the kinematic regime}

Using the meridional velocity inverted from the sunspot number time series
(see Fig.~\ref{fig:vp_c1_var}), \cite{2009SoPh..257....1L}  showed that most of the long term variability of the sunspot number could be explained as being driven by the meridional velocity decadal
variations, assuming that the evolution of the solar magnetic field
is well described by an axisymmetric kinematic dynamo model.

Figure (\ref{fig:LODM_Surya_comp}) shows a reconstructed sunspot times series that has
been obtained using the meridional velocity $v_p$ inverted from the sunspot observational
time series, and Figure (\ref{fig:compsuryssn}) shows the phase space
of a standard axisymmetric kinematic dynamo model (with the same $v_p$ for all cycles)
(\cite[e.g.,][]{1995A&A...303L..29C})  and a solar dynamo model where the $v_p$
changes from cycle to cycle as inverted from the sunspot times series (\cite{2009SoPh..257....1L}).
It is quite encouraging  to find that such class of dynamo models for which the $v_p$ changes overtime
successfully reproduced the main features found in the observational data.

Moreover, in their article~\cite{2009SoPh..257....1L} tested two different methods of implementing
the velocity variation for each magnetic cycle, namely, by considering that amplitude variations
in $v_p$ that take place at sunspot minima or at sunspot maxima.
All the time series show a few characteristics that are consistent with
the observed sunspot records. In particular,  all the simulations show
the existence of low amplitudes on the sunspot number time series
between 1800 and 1840 and  between 1870 and 1900.
The simulation that best reproduces the solar data corresponds
to the model  SSNrec[3] (see Fig.~\ref{fig:LODM_Surya_comp}), in which was implemented a smoothed
$v_p$  variation  profile between consecutive cycles and taking place at the  solar maximum.
This clearly highlights the potential of such methodology.

Here, we discuss the same methodology as the one used in the previous section,
but  instead of applying it to observational sunspot number records,
it is used to reconstruct the sunspot time series. The results obtained clearly
show that the present kinematic dynamo models can reproduce in some detail
the observed variability of the solar magnetic cycle. The fact that for one
of the sunspot models -- model SSNrec[3], it presents a strong level of correlation with the observational
time series, lead us to believe that the main idea behind this {\it backward approach}
is correct and it is very likely that the inverted $v_p$ variation is probably very close to the $v_p$
variation  that happens in the real Sun. Clearly, under the assumed theoretical framework the meridional
circulation is the leading quantity responsible for the magnetic variability found in the sunspot number
time series and current solar dynamo models are able to reproduce such variability to a certain degree.

\subsection{Using a LODM in the non kinematic regime}
\label{sec:NonKinematicLODM}

So far, the vast majority of the LODM applications presented here followed the traditional assumption that the solar dynamo can be correctly modeled in the kinematic regime, where only the plasma flows influence the production of magnetic field, and not the other way around. This kinematic approximation is used in the vast majority of the present 2.5D spatially resolved dynamo models.

In the last couple of years though, evidence started to appear supporting the claim that this kinematic regime might be overlooking important physical mechanisms for the evolution of the dynamo. The idea that the meridional flow strength can change over time and affect the solar cycle amplitude coupled with the measurements of~\cite{HathawayScience2010} and~\cite{BasuAntia2010} indicate that the observed variation in this flow is highly correlated with the levels of magnetic activity. This leads to the fundamental question: \textit{"Is the flow driving the field or is the field driving the flow?"}

The first clues are starting to appear from 3D MHD simulations of solar convection. The recent analysis
of the output of one of the large-eddy global MHD simulations of the solar convection zone
done by \cite{2012SoPh..279....1P} shows interesting clues.
These simulations solve the full set of MHD equations in the anelastic regime, in a broad,
thermally-forced stratified plasma spherical shell mimicking the SCZ and are fully dynamical on all spatiotemporally-resolved scales. This means that a two way interaction between field and flow is always
 present during the simulation.
The analysis shows that the interaction between the toroidal magnetic field and the meridional flow
in the base of the convection zone indicates that the magnetic field is indeed acting on the equatorward deep section of this flow, accelerating it. This observed relationship runs contrary to the usually assumed
kinematic approximation.

In order to check if this non-kinematic regime has any impact in the long term dynamics of the solar
dynamo, \cite{2012SoPh..279....1P} implemented a term that accounts for the Lorentz force feedback in
a LODM similar to the one presented here. This allows to fully isolate the global aspects of the
dynamical interactions between the meridional flow and magnetic field in a simplified way.

\smallskip

They assumed that the large-scale meridional circulation, $v_p$, is divided into a
``kinematic'' constant part, $v_0$ (due to angular momentum distribution) and a time dependent
part, $v(t)$, that encompasses the Lorentz feedback of the magnetic field.
Therefore they redefine $v_p$ as $v_p(t) = v_0 + v(t)$ where the time dependent part evolves
according to
\begin{equation}
    \frac{\textrm{d} v(t)}{\textrm{d} t} = a\, B_\phi A_p - b\, v(t)\,\,\,.
    \label{eq:vp}
\end{equation}

The first term is a magnetic nonlinearity representing the Lorentz force
and the second is a "newtonian drag" that mimics the natural resistance of the flow to an
outside kinematic perturbation. Under these conditions the Lorentz force associated with the cyclic
large-scale magnetic field acts as a perturbation on the otherwise dominant kinematic meridional flow.
This idea was not new and it was used before in the context of magnetically-mediated variations of
differential rotation in mean-field dynamo models \cite{Tobias1996}, \cite{MossBrooke2000} and \cite{Bushby2006}. The modified LODM equation they end up defining are

\begin{eqnarray}
    \frac{\textrm{d} B_\phi}{\textrm{d} t} &=& \left(c_{1}
    -\frac{v_p(t)}{\ell_0}\right) B_\phi
    + c_{2} A_p - c_{3} B_\phi^3~,
    \label{eq:newlodmB} \\
    \frac{\textrm{d} A_p}{\textrm{d} t} &=& \left(c_{1}
    -\frac{v_p(t)}{\ell_0}\right) A_p + \alpha  B_\phi~,
    \label{eq:newlodmA}
\end{eqnarray}
where $c_1$, is defined as $c_{1}=\frac{\eta}{\ell_0^2}-\frac{\eta}{R_\odot^2}$ and takes the role
of magnetic diffusivity, while the other coefficients remain the same.

\begin{figure}[htb!]
    \centering
    \includegraphics[width=38mm]{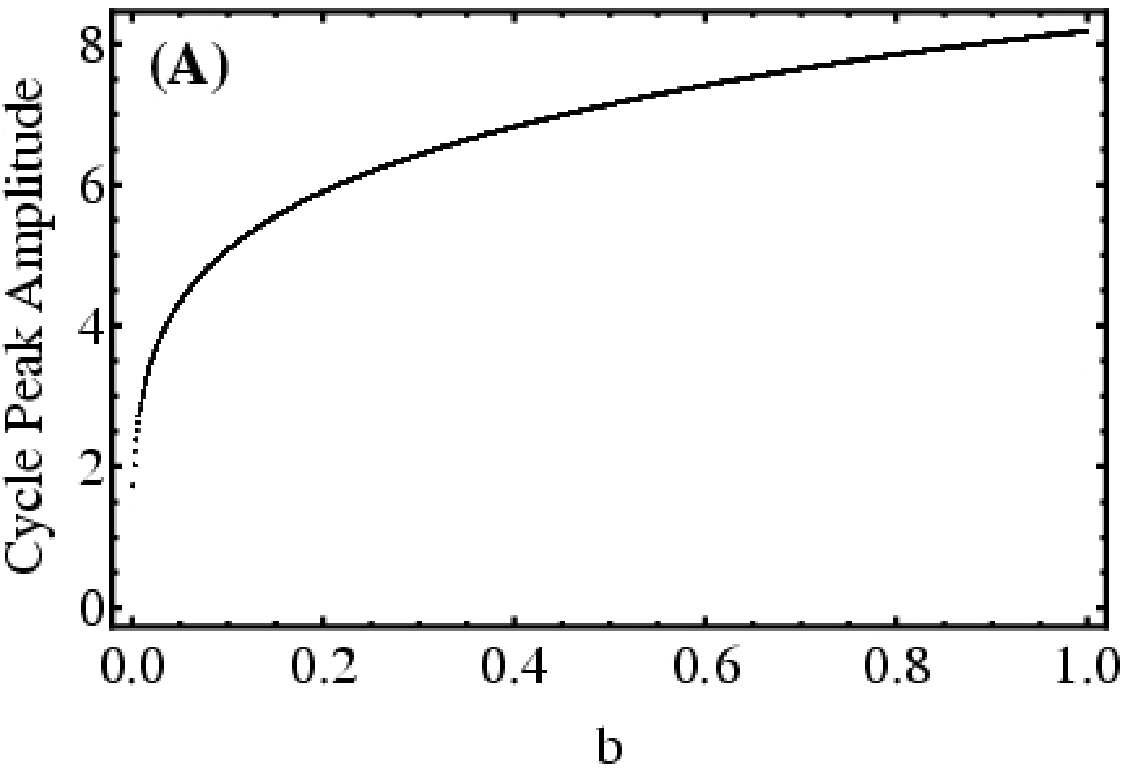}
    \includegraphics[width=38mm]{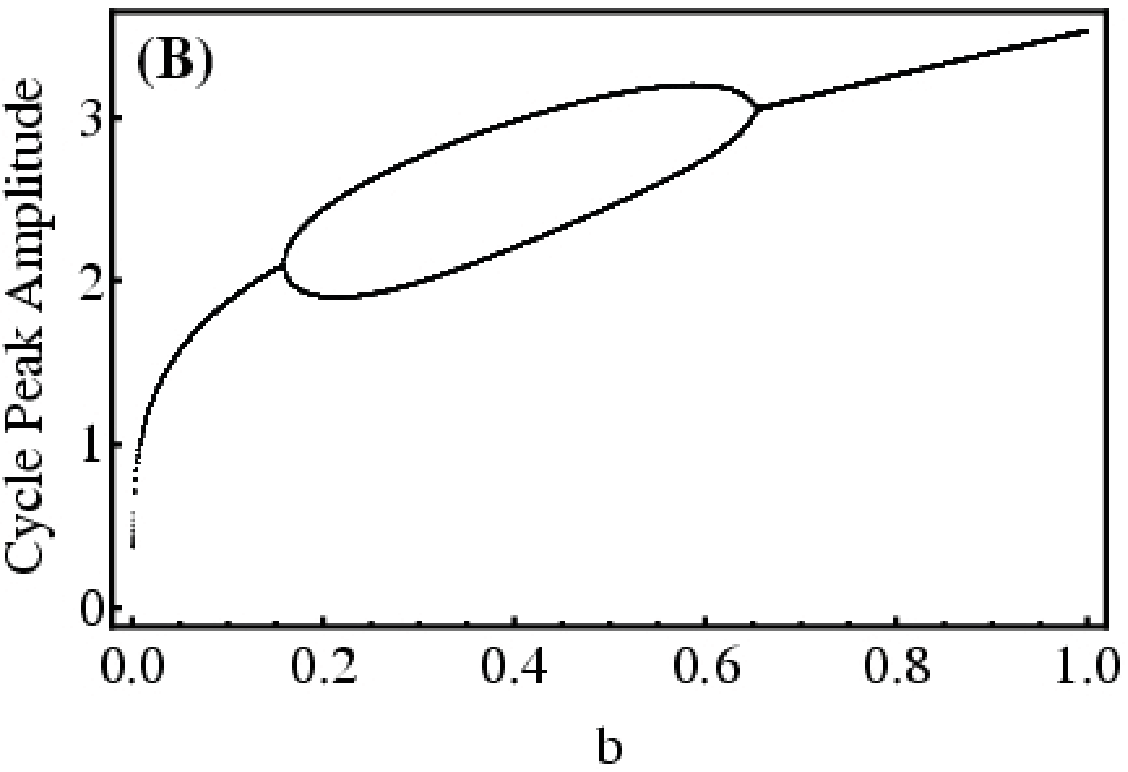}
    \includegraphics[width=38mm]{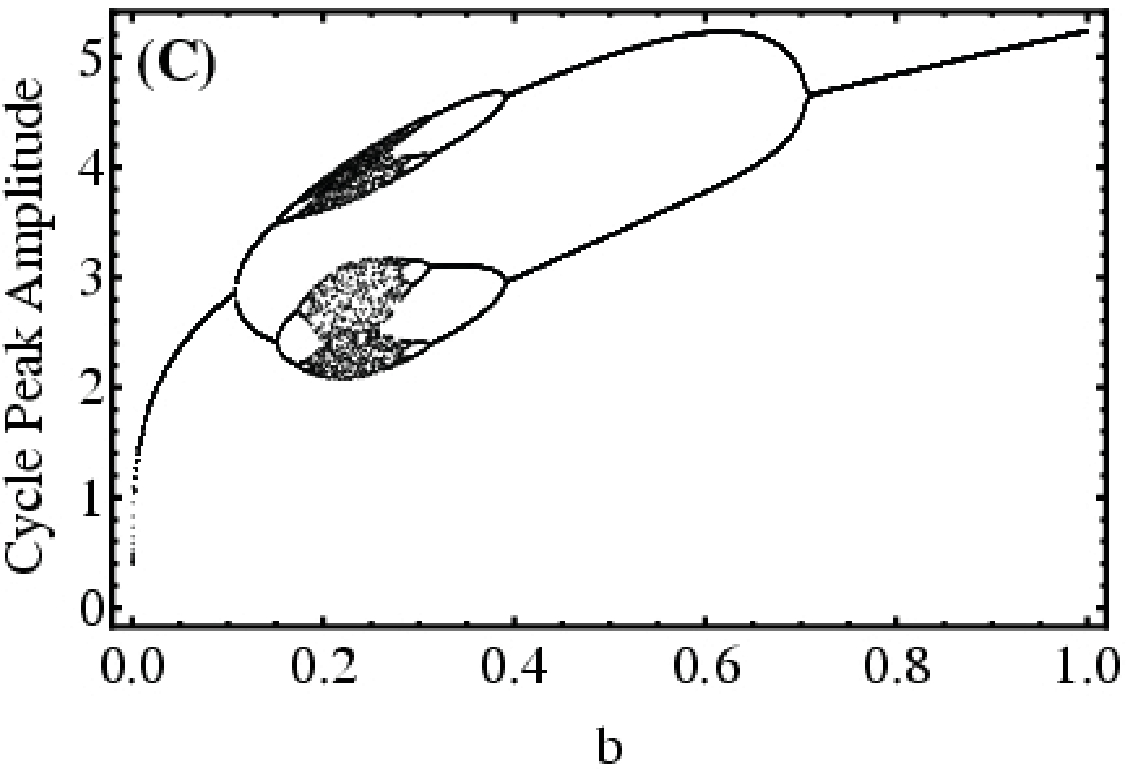}
    \caption{Bifurcation maps for maximum amplitude of the toroidal field (equivalent
    to solar cycle maximum) obtained by varying $b$ between $10^{-4}$ and $1$ for
    different $a$ and $v_0$. (A) Single period regime, $v_0=-0.1$, $a=0.01$;
    (B) Appearance of period doubling, $v_0=-0.1$, $a=0.1$ and (C) shows signatures
    of  chaotic regimes with multiple attractors and windows, obtained
    with $v_0=-0.13$, $a=0.05$. \textit{Adapted from} \cite{2012SoPh..279....1P}.}
    \label{fig-bifurcation_maps}
\end{figure}

While the values used for the structural coefficients,
are mean values extracted from the works presented in the previous in sections, the parameters associated with the meridional flow evolution, $a$, $b$ and $v_0$ deserved now the attention. These parameters have an important role in the evolution of the solution space. The behavior observed in the solutions range from fixed-amplitude oscillations closely resembling kinematic solutions, multiperiodic solutions, and even chaotic solutions.
This is easier to visualize in Figure (\ref{fig-bifurcation_maps}) where are presented analogs of classical bifurcation diagrams by plotting successive peak values of cycle amplitudes, for solutions with fixed ($a$, $v_0$) combinations but spanning through values of $b$. Transitions to chaos through bifurcations are also observed when holding $b$ fixed and varying $a$ instead.

\bigskip

\begin{figure}[htb!]
    \centering
    \includegraphics[width=100mm]{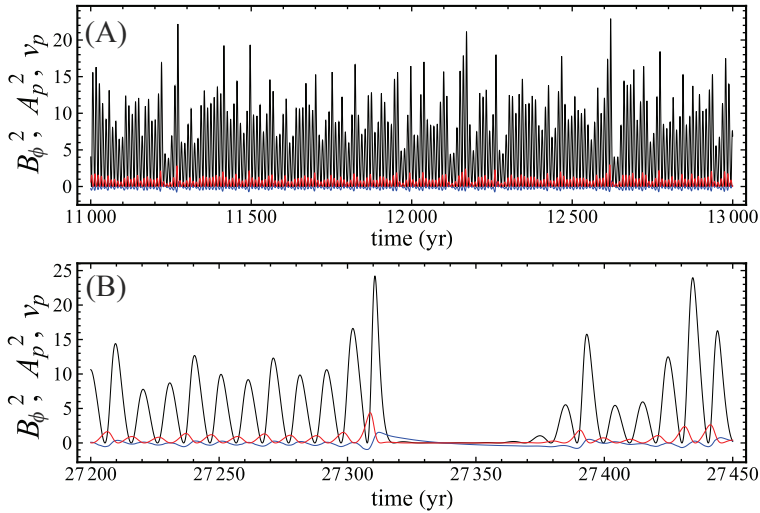}
    \caption{Simulation result fluctuating $a\in[0.01,0.03]$, $b=0.05$ and $v_0=-0.11$.
    All other model parameters are the same as in the reference solution. Panel (A) shows a
    section of the simulation where the long term modulation can be seen.
    In black is $B_\phi^2(t)$, red $A_p^2(t)$ and blue a scaled version of the meridional
    flow, in this case 5$v_p(t)$. In panel (B) the same quantities but this time zooming
    in into a grand minimum (off phase) period. \textit{Adapted from}
    \cite{2012SoPh..279....1P}.}
    \label{fig-LODM_stochastic_a}
\end{figure}

The authors expanded the methodology used and applied stochastic fluctuations to parameter $a$, the one that controls the influence of the Lorentz force.
As a result, and depending on the range of fluctuations, they observed that the short term stochastic kicks
in the Lorentz force amplitude create long term modulations in the amplitude of the cycles (hundreds of years)
 and even episodes where the field decays to near zero values, analog to the previously mentioned grand
 minima. The duration and frequency of these long quiescent phases, where the magnetic field
decays to very low values, is determined by the level of fluctuations of $a$ and the
value of $b$. The stronger this drag term $b$ is, the shorter the minima are and the higher
the level of fluctuation of $a$, the more common these intermittency episodes become.
Figure (\ref{fig-LODM_stochastic_a}) shows a section of a solution that spanned for 40000 years
and that presents all the behaviors described before.

In this specific example they used 100\% fluctuation in $a$ and maintaining all the other
parameters constant. In the the parameter space used to produce this figure, the solution
without stochastic forcing is well behaved in the sense that it presents a single period
regime. Therefore, the fluctuations observed in this solution are a direct consequence
of the stochastic forcing of the Lorentz force and not from a chaotic regime of the
solution's space.

\smallskip

To understand how the grand minima episodes arise they resort to
  visualizing one of these episodes with phase space diagrams of
\{$B_\phi$, $A_p$, $v_p$\}. This allows to see how these quantities vary in relation to
each other and try to understand the chain of events that trigger a grand minimum.

\begin{figure}[htb!]
    \centering
 \includegraphics[width=100mm]{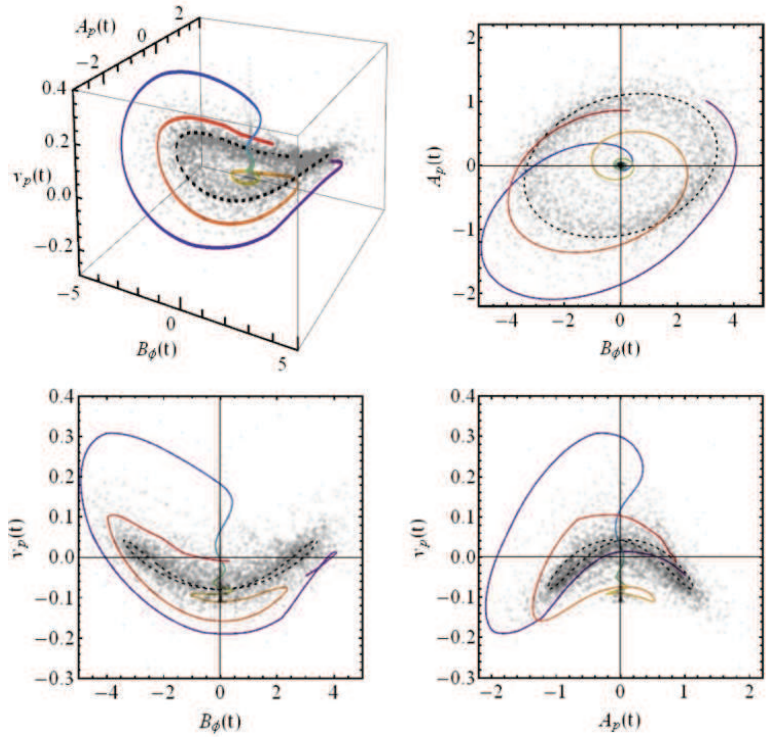}
\caption{\small Phase spaces of the solution with stochastic fluctuations. The gray
dots represent 1 year intervals between t=35000 and t=40000. The colored line shows
the trajectory of a grand minima (starting from purple, t=27300 and ending in
red, t=27400. The black dashed line represents the unperturbed solution with $a=0.02$.
\textit{Adapted from} \citet{2012SoPh..279....1P}.}
\label{fig-LODM_phasespace}
\end{figure}

The standard solution for the LODM without stochastic forcing, i.e. with $a$ fixed at
the mean value of the random number distribution used, is the limit cycle attractor, i.e.,
a closed trajectory in the \{$B_\phi$, $A_p$\} phase space. This curve is represented
as a black dashed trajectory in the panels of Figure (\ref{fig-LODM_phasespace}).
The gray points in this figure are the stochastic forced solution values sampled at
1 year interval.
These points scatter around the attractor representing the variations in amplitude of
the solution. Occasionally the trajectories defined by these points collapse
to the center of the phase space (the point \{0,0,$v_0$\} is also another natural
attractor of the system) indicating a decrease in amplitude of the cycle, i.e. a grand minimum.
The colored trajectory evolving in time from purple to red represents one of those grand minimum.
This happens when the solution is at a critical distance from the limit cycle attractor
and gets a random kick further away from it. This kick makes the field grow rapidly.
In turn, since the amplitude of the field grows fast, the Lorentz force will induce
a similar growth in $v(t)$ eventually making $v_p$ change sign. When this occurs, $v_p$ behaves
as a sink term quenching the field growth very efficiently.
This behavior is seen in the two bottom panels of Figure (\ref{fig-LODM_phasespace})
where $v_p$ decays to its imposed ''kinematic'' value
$v_0$ after the fields decay. After this collapse of $v_p$ to $v_0$ it starts
behaving has a source term again and the cyclic activity proceeds.

\bigskip

One clear advantage of low order models emerges from this example. Currently 3D MHD simulations of solar convection spanning
a thousand years take a couple of months to run in high efficiency computational clusters or in supercomputers.
Longer simulations are at the moment prohibitive not only for the amount of time they take but also for the huge amount of
data they generate. Statistical studies on grand minima originated by the kind of magnetic back-reaction described here,
require long integration times where many thousands of cycles need to be simulated. The LODM calculations can be done in a
few minutes or hours in any current desktop.

The grand minima mechanism presented in this section is now being studied by looking at the data
available from 3D simulations. Some effects are easier to find when you know what to look for.

\section{Outlook for the Sun and Stars}
\label{sec:Outlook}
\begin{figure}
\centering
\includegraphics[scale=0.65]{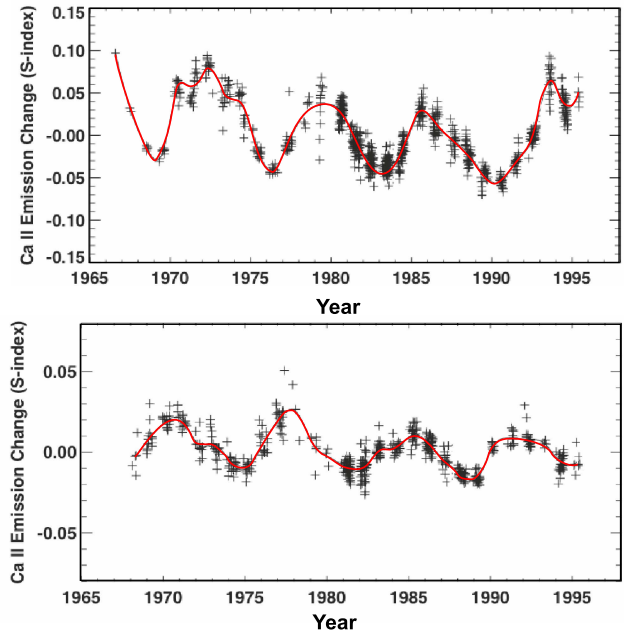}
\caption{Magnetic activity signature expressed by the variation of the intensity of the
Ca II emission line (S-index) for two solar like stars. \textit{Adapted from}~\citet{1995ApJ...438..269B}.}
\label{fig:msstar}
\end{figure}

So far we have shown that low order dynamo models (for which
the approximation must be carefully chosen to keep the relevant
physics within)  could lead the way to explore some features of
the solar magnetic activity including the long-term variability.  The
study of the phase diagram  $\{B_\phi(r),\dot{B}_\phi (t)\}$ clearly
shows that on a scale of a few centuries the solar magnetic cycle
shows evidence for a van der Pool attractor --- put in evidence by the
mean solar magnetic cycle, although on a time-scale of a few solar
magnetic cycles the phase space trajectory changes dramatically.  In
some cases the trajectory collapses completely for several magnetic
cycles as in the periods of grand minima. This gives
us an indication about the existence of a well defined self regulated
system under all this  observed magnetic variability, for which we
still need to identify the leading physical mechanisms driving the solar
dynamo to extreme activity scenarios like periods of grand minima.
Actually, the fact that a well-defined averaged van der Pool limit
curve exists for all the sunspot records, can be used  to test
different solar dynamo models, including numerical simulations,
against observational data or between different dynamo models.

Moreover, the fact that such well-defined attractor exists in the phase space,
and several dynamo models are able to qualitatively reproduce the solar variability
(as observed in the phase space $\{B_\phi(r),\dot{B}_\phi (t)\}$
gives us hope that in the near future we will be able to make quite reliable
short term predictions of the solar magnetic cycle variability,
at least within certain time intervals of solar magnetic activity.
A significant contribution can be done by the utilization of more
accurate sunspot time series in which
many of the historical inaccuracies were corrected~\citep{2014SoPh..289..545L}.

In the future, similar inversion techniques could be developed, namely to study
the possible asymmetry between the North and South Hemispheres using the sunspot areas,
either by treating each of the sunspot areas as two distinct times series or by
attempting two-dimension inversions of sunspot butterfly diagrams.
In the former case, recently \citet{Lopesetal2014} have analysed 
these long-term sunspot areas time series and found that turbulent convection  
and  solar granulation are responsible  by the stochastic nature of the sunspot area variations.
In the last case, we could learn about the evolution of the solar magnetic cycle
in the tachocline during the last two and a half centuries.
Moreover, most of the inversion methods used for the sunspot number can be easily
extended to other solar magnetic cycles proxies such as TSI, H$_\alpha$ and Magnetograms.

The oscillator models, as a first order dynamo model are
particularly suitable to study the magnetic activity in other stars. A
good proxy of  magnetic activity in stars in the chromospheric
variations of Ca II H and K emission lines.
\citet{1995ApJ...438..269B} have found many F2 and M2 stars which seem
to have cyclic magnetic cycle activity, as observed in the Sun
(see Fig.~\ref{fig:msstar}). In some of these stars the observational time
series covers several cycles of activity.  In particular, it will be
interesting to identify how the dynamo operating in these stars
differs from the solar case.

More recently, the CoROT and Kepler space missions have observed photometric
variability associated with solar-like activity in a very large number
of main sequence and sub-giant stars. While the time coverage is too short to derive cycle
periods for stars very close to the Sun, the overall {\it level} of activity
and its dependence on various stellar parameters can be studied on a large statistical
sample \citep{Basri2010,McQuillan2012}. Nevertheless, with so many
stars with quite distinct masses and radius, it is reasonable to expect
that we will find quite different type of dynamos and regimes of stellar magnetic cycle.
Actually, we think it is likely to find a magnetic diversity identical
to the one found in the acoustic oscillation spectra measured for the more
than 500 sun-like stars already discovered~\citep{2014ApJS..210....1C},
some of which have already shown evidence of a magnetic cycle activity.
\citet{2010Sci...329.1032G} have obtained a proxy of the starspots number for the star HD49933
from amplitudes and frequencies of the acoustic modes of vibration.
As in the nonlinear oscillator models the activity level is determined by
the structural parameters which in turn depends on the dynamo model.
These studies potentially offer a simple theoretical scheme against
which to test the observational findings.

\begin{acknowledgements}
The authors thank the anonymous referee for the suggestions made to improve the quality of the article.
I.L. thanks the convenors of the workshop and The International Space Science Institute  
for the invitation and financial support. 
I.L. and D.P. would also like to thank Arnab Choudhuri and his collaborators for making the Surya code publicly available.
I.L. would like to thank his collaborators in this subject of research: Ana
Brito, Elisa Cardoso, Hugo Silva, Amaro Rica da Silva and Sylvaine Turck-Chi\`eze.
The work of I.L. was supported by grants from "Funda\c c\~ao para a Ci\^encia e Tecnologia"
and "Funda\c c\~ao Calouste Gulbenkian". D.P. acknowledges the
support from the Funda\c{c}\~{a}o para a Ci\^{e}ncia e Tecnologia (FCT) grant
SFRH/BPD/68409/2010. M. N. acknowledges support from the
Hungarian Science Research Fund (OTKA grant no.\ K83133). 
\end{acknowledgements}

\newpage


\end{document}